\documentclass[12pt]{article}
\usepackage{graphicx}
\usepackage{epsfig}
\usepackage{epstopdf}
\DeclareGraphicsExtensions{.pdf,.eps,.png,.jpg,.mps}

\def\lsim{\mathrel{\rlap{\lower3pt\hbox{\hskip0pt$\sim$}}
     \raise1pt\hbox{$<$}}}         
\def\gsim{\mathrel{\rlap{\lower4pt\hbox{\hskip1pt$\sim$}}
     \raise1pt\hbox{$>$}}}         

\usepackage{amsmath}
\usepackage{amsfonts}

\begin{document}
\begin{titlepage}

\centerline{\Large \bf Volatility Smile as Relativistic Effect}
\medskip

\centerline{Zura Kakushadze$^\S$$^\dag$\footnote{\, Zura Kakushadze, Ph.D., is the President of Quantigic$^\circledR$ Solutions LLC, and a Full Professor at Free University of Tbilisi. Email: \tt zura@quantigic.com}}
\bigskip

\centerline{\em $^\S$ Quantigic$^\circledR$ Solutions LLC}
\centerline{\em 1127 High Ridge Road \#135, Stamford, CT 06905\,\,\footnote{\, DISCLAIMER: This address is used by the corresponding author for no
purpose other than to indicate his professional affiliation as is customary in
publications. In particular, the contents of this paper
are not intended as an investment, legal, tax or any other such advice,
and in no way represent views of Quantigic® Solutions LLC,
the website \underline{www.quantigic.com} or any of their other affiliates.
}}
\centerline{\em $^\dag$ Free University of Tbilisi, Business School \& School of Physics}
\centerline{\em 240, David Agmashenebeli Alley, Tbilisi, 0159, Georgia}
\medskip
\centerline{(August 20, 2016; revised January 29, 2017)}

\bigskip
\medskip

\begin{abstract}
{}We give an explicit formula for the probability distribution based on a relativistic extension of Brownian motion. The distribution 1) is properly normalized and 2) obeys the tower law (semigroup property), so we can construct martingales and self-financing hedging strategies and price claims (options). This model is a 1-constant-parameter extension of the Black-Scholes-Merton model. The new parameter is the analog of the speed of light in Special Relativity. However, in the financial context there is no ``speed limit" and the new parameter has the meaning of a characteristic diffusion speed at which relativistic effects become important and lead to a much softer asymptotic behavior, i.e., fat tails, giving rise to volatility smiles. We argue that a nonlocal stochastic description of such (L{\'e}vy) processes is inadequate and discuss a local description from physics. The presentation is intended to be pedagogical.
\end{abstract}
\medskip
\end{titlepage}

\newpage

\section{Introduction}

{}A host of mathematical methods for describing physical phenomena have been gobbled up into quantitative finance. One of the most vivid examples is Brownian motion. Methods of classical mechanics, statistical physics, and even quantum mechanics (see below), are widely used in quantitative finance. One area of physics that -- to a large extent -- has eluded financial applications is Einstein's Special Relativity.

{}There are two reasons. First, in most aspects of finance (e.g., asset pricing) the finiteness of the speed of light is utterly irrelevant, so any direct application of Special Relativity would be farfetched at best.\footnote{\, Albeit, it has in fact been argued in the context of high frequency trading (HFT) that relativistic effects literally may become relevant therefor; see, e.g.,
\cite{Wissner},
\cite{Laughlin},
\cite{Buchanan}.
These arguments are essentially anchored on a simple observation that, since the speed of light is finite, it can take enough time for a signal to propagate from substantially spatially separated points (locations of the servers involved in trading) on Earth potentially to make an arbitrageable difference. We will not delve into this train of thought here. For other works alluding to relativistic effects, see, e.g.,
\cite{Haug},
\cite{Tenreiro},
\cite{Romero}.
We comment on the last reference at the end of Section \ref{sec.5}.\label{HFT}} Second, in Special Relativity spacetime has a Minkowski signature, e.g., in 2-dimensional Minkowski spacetime (with one spatial $x$ and one time $t$ direction) the distance between two points is given by $s^2 =(x_1 - x_2)^2 - (t_1 - t_2)^2$. However, Minkowski spacetime is utterly inapplicable in finance, at least in the context of asset pricing. Thus, if we take Brownian motion, it is equivalent to a free quantum mechanical particle in Euclidean time, not in usual (Minkowski) time. Euclidean time $t_E$ is related to Minkowski time $t_M$ via the so-called Wick rotation $t_M = -i t_E$ \cite{Wick}, i.e., in finance we are in imaginary time as far as physics is concerned. Therefore, 2-dimensional spacetime that might be relevant in the financial context would have to have a Euclidean signature with the distance given by $s^2 =(x_1 - x_2)^2 + (t_1 - t_2)^2$. And this makes all the difference.

{}For relativistic effects to be relevant in finance, there is no need for them to be literally related to (the finiteness of) the speed of light -- the latter pertains to Minkowski spacetime and is not of much relevance in finance (cf.  fn. \ref{HFT}). Instead, relativistic effects in Euclidean spacetime have nothing to do with the speed of light (as a constant of nature), but, e.g., with a characteristic speed of diffusion (in the Brownian motion lingo) at which higher-order (in powers of the diffusion speed) effects can no longer be neglected. A priori these higher-order effects can be whatever we wish them to be. However, we can constrain them by requiring that they are of the same form as analogous effects in Special Relativity in Minkowski spacetime. Practically speaking, we can start with a relativistic theory in Minkowski time and Wick-rotate it to Euclidean time. As we will see below, this almost gives us the right answer, but not quite. And the right answer is rather intriguing...

{}Our motivation and context for doing all this are volatility smiles in option prices. They are artifacts of using Gaussian probability distributions as in the Black-Scholes-Merton (BSM) model \cite{BS}, \cite{Merton}. Real-life distributions are fat-tailed and result in volatility smiles (see, e.g., \cite{Hull}). There are various extensions of the BSM model such as stochastic volatility\footnote{\, For a partial list, see
\cite{Cox},
\cite{Merton1},
\cite{CoxRoss},
\cite{CIR},
\cite{HW},
\cite{Wiggins},
\cite{Heston},
\cite{Engle},
\cite{Duan},
\cite{Chen},
\cite{Duffie},
\cite{Fouque},
\cite{Geman},
\cite{Hagan},
\cite{Kou},
\cite{Linetsky},
\cite{CarrSun},
and also references contained therein.} and local volatility\footnote{\, See, e.g.,
\cite{Dupire}, \cite{Derman}, \cite{Derman3}, \cite{Dumas}, \cite{Derman2}, \cite{Mercurio,Mercurio2}, \cite{Gatheral}, and references therein.} models, which lead to fat tails. As it turns out, relativistic effects also give rise to fat tails. The aforesaid higher-order effects soften the decay of the probability distribution at large $x$, hence fat tails. So, the idea is to make Brownian motion relativistic in Euclidean (not Minkowski) time. The analog of the speed of light then is a characteristic diffusion speed at which relativistic effects are sizable.

{}This is precisely the approach we follow here. Most of the literature on relativistic Brownian motion has been focused on Minkowski spacetime.\footnote{\, For physics related works, see, for instance,
\cite{Hakim},
\cite{Rylov},
\cite{Ben},
\cite{Sornette1},
\cite{Debbasch1997},
\cite{Debbasch1998},
\cite{Dunkel2005a, Dunkel2005b, Dunkel2009},
\cite{Oron},
\cite{Fa},
\cite{Felix}, and references therein.}
Even the telegraph equation, which arises in relativistic heat conduction,\footnote{\, For relativistic heat conduction, see, e.g., \cite{Vernotte}, \cite{Cattaneo}, \cite{Chester}.} is in Minkowski time and is not applicable for our purposes here.\footnote{\, More precisely, the telegraph equation has a ``wrong-signed" (for our purposes here, that is) second-order time derivative. In fact, as we discuss below, second-order (in time derivatives) equations are not suitable for pricing purposes. There should be only a single time derivative.} So, we discuss how to construct the probability distribution for relativistic Brownian motion in Euclidean time. A natural way to proceed is to start from the action for a 1-dimensional relativistic point particle in Minkowski time and Wick-rotate it to Euclidean time. The probability distribution then can be obtained using path integral techniques by writing the action in a reparametrization invariant form, fixing the gauge and reducing the path integral to a Gaussian (i.e., calculable) form,\footnote{\, Not to be confused with a ``Gaussian approximation" mentioned below. This trick is exact.} which can be readily evaluated. This would likely be a high energy theorist's approach. However, if we (naturally) assume the path integral measure to be the simplest possible one, then we run into two problems. First, while the resultant probability distribution is expressly invariant under $SO(2)$ rotations in 2-dimensional Euclidean spacetime, it does not normalize to 1 when we integrate over all positions $x$. Instead, the normalization decays exponentially in time and -- in the financial context -- corresponds to a fixed constant interest rate. Its physical interpretation is Einstein's famous rest energy $E_0=mc^2$, where $m$ is the mass of the particle and $c$ is the speed of light.\footnote{\, The aforesaid constant interest rate is then given by $r_0 = E_0/\hbar$ ($\hbar$ is the reduced Planck constant) and the map to finance is via identifying $\hbar/m$ with $\sigma^2$ ($\sigma$ is the diffusion coefficient -- see below). The financial interpretation of $c$ is given below. It is {\em not} a maximal speed (cf. physics).} We must remove this contribution at the expense of giving up the aforesaid $SO(2)$ rotational invariance, which is just as well as it is clearly absent in the context of finance. Second, even ignoring the rest energy issue, the probability distribution does not obey the tower law (a.k.a. semigroup property), which is a serious issue as it is utterly unclear how to construct martingales and price options and other claims in the absence thereof (see below).

{}We should emphasize that it is not that the aforesaid path integral computation is wrong. It is just that the na\"{i}ve (i.e, the simplest) choice of the path integral measure is not suitable for our intended financial applications. Happily, there is a rather simple way to circumvent this issue. We can use the Hamiltonian approach, where the probability distribution obeys a first-order (in time derivatives) differential equation. The price to pay is that in the relativistic case the Hamiltonian is a non-polynomial (i.e., nonlocal) differential operator w.r.t. $x$ derivatives. However, a priori this does not pose an issue and, by requiring i) proper normalization and ii) the tower law, we can uniquely determine the probability distribution:\footnote{\, Sometimes referred to as the normal inverse Gaussian distribution \cite{BN}. It was first derived in physics by \cite{Newton}. The $K_\alpha(\cdot)$ function also appears in the variance gamma models, see, e.g., \cite{Madan}. However, there $\alpha$ is time-dependent.}
\begin{equation}\label{Bessel}
 {\widetilde P}(x, t) = {c\over\pi\sigma^2} {ct\over\sqrt{x^2 + c^2t^2}} K_1\left({c\over\sigma^2}\sqrt{x^2 + c^2 t^2}\right)\exp(c^2 t/\sigma^2)
\end{equation}
Here $\sigma$ is what we would interpret as the volatility (i.e., the diffusion coefficient) in the non-relativistic case, so $\sigma^2$ (which has dimension of $1/t$ assuming $x$ is dimensionless) is the diffusion speed, while $c$ is the analog of the speed of light and in the financial context is interpreted as a characteristic diffusion speed at which relativistic effects become important. The non-relativistic limit, i.e., the Gaussian distribution, is (formally) recovered in the $c\rightarrow\infty$ limit.\footnote{\, Via the asymptotic property of modified Bessel functions $K_\alpha(x)\sim\sqrt{\pi/2x}\exp(-x)$ at $x\gg 1$.} However, this limit is misleading. Thus, the large $|x|$ behavior of (\ref{Bessel}) is given by
\begin{equation}\label{fat.tails}
 {\widetilde P}(x, t) \approx {1\over\sqrt{2\pi\sigma^2 t}} \left[1 + {x^2\over c^2t^2}\right]^{-{3\over 4}} \exp\left(-{c^2t\over\sigma^2}\left[\sqrt{1 + {x^2\over c^2 t^2}} - 1\right]\right)
\end{equation}
So, the decay at large $|x|$ is much softer and we have fat tails.\footnote{\, The Laplace distribution (exponential decay) arises in \cite{CarrLap}. Ours is not Laplacian.} The moral of the story is that a ``Gaussian approximation" (in this particular case, the $c\rightarrow \infty$ limit) is not as innocuous as it might appear.\footnote{\, In Section \ref{sec.5} we give an example of this from physics. It is also related to expanding $\sqrt{1+y^2}$.} Put differently, it is simply inadequate.

{}The remainder hereof is organized as follows. In Section \ref{sec.2} we quickly review a Euclidean non-relativistic quantum particle, which is Brownian motion. In Section \ref{sec.3} we connect Euclidean path integral to asset pricing.\footnote{\, Sections \ref{sec.2} and \ref{sec.3} somewhat overlap with \cite{ShortRate} (the path integral approach).} In Section \ref{sec.4} we port this discussion over to the relativistic case. We use path integral to compute the aforesaid $SO(2)$ invariant probability distribution, point out its shortcomings, discuss how to fix them using the Hamiltonian approach and arrive at the correct probability distribution by requiring proper normalization and the tower law. We discuss martingales (and an upper bound on the log-volatility not present in the BSM model), option pricing (via self-financing hedging strategies), and volatility smiles due to fat tails. We argue that a nonlocal stochastic description of such (L{\'e}vy) processes is inadequate and discuss a local description from physics. We conclude in Section \ref{sec.5}.

\section{Non-Relativistic Mechanics}\label{sec.2}

{}Newton's second law for one-dimensional motion along the $x$-axis
\begin{equation}\label{N2}
 F = m~a
\end{equation}
where $m$ is the mass of an object, $a = \ddot{x}$ is its acceleration,\footnote{\, Each dot over a variable stands for a time derivative.}  and $F$ is the force acting on it, can be derived via the principle of stationary action. Let $x(t)$ be a continuous path ${\cal F}$ connecting two spacetime points $(x_0, t_0)$ and $(x_f, t_f)$, where $t_f > t_0$. The action functional $S = S[x(t)]$ is defined as
\begin{equation}\label{action}
 S = \int_{\cal F} dt~L(x, \dot{x}, t)
\end{equation}
where $L$ is the Lagrangian:
\begin{equation}\label{nr.lag}
 L = E_K - V
\end{equation}
Here
\begin{equation}\label{kinetic}
 E_K = {1 \over 2} mv^2
\end{equation}
is the kinetic energy of the object ($v = \dot{x}$ is the velocity), and the function $V = V(x, t)$ is called potential energy. Now consider a small variation of the path $x(t) \rightarrow x(t) + \delta x(t)$, where $\delta x(t)$ vanishes at the endpoints of the path: $\delta x(t_0) = \delta x(t_f) = 0$. The variation of the action reads:
\begin{equation}
 \delta S = \int_{\cal F} dt~\left[{\partial L\over\partial x} - {d\over dt}{\partial L\over\partial \dot{x}}\right]\delta x(t)
\end{equation}
where we have integrated by parts and taken into account that the surface term vanishes. The functional derivative $\delta S/\delta x(t)$ vanishes if and only if\footnote{\, This is because $\delta x(t)$ is arbitrary subject to the boundary conditions at $t_0$ and $t_f$.}
\begin{equation}\label{EL}
 {d\over dt}{\partial L\over\partial \dot{x}} = {\partial L\over\partial x}
\end{equation}
This is the Euler-Lagrange equation, which is the equation of motion (\ref{N2}) with the identification $F = -{\partial V/ \partial x}$.
Classical trajectories are deterministic: (\ref{N2}) is a second-order differential equation, so the path $x(t)$ is uniquely fixed by specifying the endpoints $(x_0, t_0)$ and $(x_f, t_f)$. Alternatively, it is uniquely determined by specifying the initial conditions: $x(t_0) = x_0$ and $\dot{x}(t_0) = v_0$, where $v_0 = v(t_0)$ is the initial velocity.

\subsection{Quantum Mechanics and Path Integral}

{}In contrast, quantum mechanics is not deterministic but probabilistic. One can only determine the probability $P(x_0, t_0; x_f, t_f)$ that starting at $(x_0, t_0)$ a quantum particle will end up at $(x_f, t_f)$. This is because of Heisenberg's uncertainty principle: in quantum mechanics it is not possible to specify both the position and the velocity at the same time with 100\% certainty.\footnote{\, See \cite{Heisenberg}, \cite{Kennard}, \cite{Weyl}.} One way to think about this is that, starting at $(x_0, t_0)$, the particle can take an infinite number of paths with a probability distribution. The probability $P(x_0, t_0; x_f, t_f)$ is given by
\begin{equation}
 P(x_0, t_0; x_f, t_f) = \left|\langle x_f, t_f\left.\right| x_0, t_0\rangle\right|^2
\end{equation}
where the probability amplitude\footnote{\, A.k.a. wave function, matrix element, propagator or correlator.} is given by Feynman's path integral (Feynman, 1948)\footnote{\, A.k.a. functional integral or infinite dimensional integral.}
\begin{equation}\label{FPI}
 \langle x_f, t_f\left.\right| x_0, t_0\rangle = \int_{x(t_0) = x_0,~x(t_f) = x_f} {\cal D}x~\exp\left({i\over\hbar}~S\right)
\end{equation}
where the integration is over all paths connecting points $(x_0, t_0)$ and $(x_f, t_f)$, and ${\cal D}x$ includes an appropriate integration measure, which we will define below. Also, $\hbar$ is the (reduced) Planck constant, which is measured experimentally.

{}The path integral (\ref{FPI}) can be thought of as an $N\rightarrow \infty$ limit of $N-1$ integrals. Let us break up the interval $[t_0, t_f]$ into $N$ subintervals: $t_i = t_{i-1} + \Delta t_i$, $i=1,\dots,N$, with $t_N = t_f$. Let $x_i = x(t_i)$, with $x_N = x_f$. We can discretize the derivative $\dot{x}(t_i)$ via $(x(t_i) - x(t_{i-1}))/\Delta t_i = (x_i - x_{i-1})/\Delta t_i$, $i=1,\dots, N$. The integral in the action (\ref{action}) can be discretized as follows\footnote{\, As usual, there are choices in defining the discretized derivative and integral, which are essentially immaterial in the continuum limit.}
\begin{equation}
 S_{N} = \sum_{i=1}^N \Delta t_i \left[{m(x_i - x_{i-1})^2\over 2\Delta t_i^2} - V(x_{i-1}, t_{i-1})\right]
\end{equation}
Then the path integral (\ref{FPI}) can be defined as
\begin{equation}\label{discrete}
 \int_{x(t_0) = x_0,~x(t_f) = x_f} {\cal D}x~\exp\left({iS\over\hbar}\right) = \lim_{N \to \infty} \prod_{i=1}^{N} \sqrt{m\over 2\pi i \hbar\Delta t_i} \int \prod_{i=1}^{N-1} dx_i~\exp\left({iS_N\over\hbar}\right)
\end{equation}
where each of the $(N-1)$ integrals\footnote{\, We have $(N-1)$ integrals as $x_N$ is fixed: $x_N = x_f$.} over $x_1,\dots,x_{N-1}$ is over the real line ${\bf R}$. We will not derive the normalization of the measure in (\ref{discrete}). In the context of quantum mechanics it can be thought of as being fixed either by using (\ref{discrete}) and (\ref{FPI}) as the {\em definition} of the probability amplitude and comparing it with the experiment, or by matching it to equivalent formulations of quantum mechanics, e.g., Schr\"odinger's equation, which itself is compared with the experiment. In the context of stochastic processes, path integral interpretation is different than in quantum mechanics and we will fix the measure directly using the definition of (conditional) expectation.

\subsection{Euclidean Path Integral}\label{sub.euclidean}

{}Mathematically, the complex phase in Feynman's path integral (\ref{FPI}) might be a bit unsettling. The complex phase disappears if we go to the so-called Euclidean time via the Wick rotation $t\to -it$. We then have Euclidean path integral:
\begin{equation}\label{EPI}
 \langle x_f, t_f\left.\right| x_0, t_0\rangle = \int_{x(t_0) = x_0,~x(t_f) = x_f} {\cal D}x~\exp\left(-{S_E\over\hbar}\right)
\end{equation}
where
\begin{equation}
 S_E = \int dt~L_E(x,\dot{x}, t)
\end{equation}
and\footnote{\, It is assumed that, if there is any explicit $t$-dependence in $V$, it is such that $V$ is real.}
\begin{equation}\label{1D.lag}
 L_E = {1\over 2} m\dot{x}^2 + V
\end{equation}
The Euler-Lagrange equation is still of the form (\ref{EL}) with $L$ replaced by $L_E$. For notational simplicity, we will drop the subscript ``$E$" as this causes no confusion.

{}Mathematically, Euclidean path integral looks more ``well-defined" than Feynman's path integral, at least for $V\geq 0$, as in this case the argument of the exponent is a real non-negative number. The discretized version is defined via
\begin{equation}\label{discrete.E}
 \int_{x(t_0) = x_0,~x(t_f) = x_f} {\cal D}x~\exp\left(-{S\over\hbar}\right) = \lim_{N \to \infty} \prod_{i=1}^N \sqrt{m\over 2\pi \hbar\Delta t_i} \int \prod_{i=1}^{N-1} dx_i~\exp\left(-{S_{N}\over\hbar}\right)
\end{equation}
where
\begin{equation}
 S_{N} = \sum_{i=1}^N \Delta t_i \left[{m(x_i - x_{i-1})^2\over 2\Delta t_i^2} + V(x_{i-1}, t_{i-1})\right]
\end{equation}
is the discretized action. Functional integration was originally applied to asset (option) pricing in \cite{Bouchaud}, and subsequently in \cite{Baaquie0}, \cite{Otto}, \cite{Kleinert2}, etc. See \cite{ShortRate} for a more detailed list.

\section{Path Integral in Asset Pricing}\label{sec.3}

{}Euclidean path integral naturally arises in asset pricing. Suppose we have a stock $S_t$ and a cash bond $B_t$. Let us assume that $B_t$ is deterministic. Suppose $X_T$ is a claim\footnote{\, E.g., this could be a call/put/binary option or some other derivative.} at maturity $T$. Then the price of the claim at time $t$ is given by
\begin{equation}\label{claim.prc}
 V_t = B_t~\mathbb{E}\left( B_T^{-1}X_T \right)_{{\bf Q},{\cal F}_t}
\end{equation}
Here ${\bf Q}$ is the measure under which the discounted stock price $Z_t = B_t^{-1}S_t$ is a martingale, and the conditional expectation $\mathbb{E}\left(\cdot\right)_{{\bf Q},{\cal F}_t}$ is defined along the latter portion of paths with initial segments ${\cal F}_t$. Below we discuss such conditional expectations in the path integral language. Our discussion is general and not limited to stocks.

{}Consider a ${\bf P}$-Brownian motion\footnote{\, A.k.a. a Wiener process.} $W_t$ between $t=0$ and some horizon time $T$ (here ${\bf P}$ is the measure).
Let $x(t)$ be the values of $W_t$ (with $x(0)=0$). We will
divide the time interval $[t_0,t_f]$, $0\leq t_0<t_f\leq T$, into $N$ subintervals
$[t_{i-1},t_i]$, $t_N = t_f$, $t_i-t_{i-1}= \Delta t_i>0$. Let $x_i = x(t_i)$, $x_N= x_f$,
$\Delta x_i=
x_i-x_{i-1}$. Let $A_t$, $0\leq t\leq T$, be a previsible process, i.e., $A_t$ depends only on the path
${\cal F}_t=\{(x(s),s)|s\in [0,t]\}$:
\begin{equation}
 A_t=A({\cal F}_t)
\end{equation}
The conditional expectation\footnote{\, Here ${\cal F}_{t_0}=\{(x_*(s),s)|s\in [0,t_0],
x_*(0)=0, x_*(t_0)=x_0\}$, where $x_*(s)$ is fixed.} $\mathbb{E}\left( A_{t_f} \right)_{{\bf P},{\cal F}_{t_0}}$ can be thought of as a $\Delta t_i\rightarrow 0$, i.e., $N\rightarrow \infty$, limit of the corresponding discrete expression:\footnote{\, In this expectation only $x(t_0)=x_0$ is fixed and $x_N = x_f$ is integrated over -- see below.}
\begin{equation}\label{discrete.Br}
 \mathbb{E}\left( A_{t_f} \right)_{{\bf P},{\cal F}_{t_0}}=
 \lim_{N\to\infty}~\prod_{i=1}^N \int_{-\infty}^\infty {dx_i\over
 \sqrt{2\pi \Delta t_i}}\exp\left(-{(\Delta x_i)^2\over 2\Delta t_i}\right)~
 A_{t_f,{\cal F}_{t_0}}
\end{equation}
where
\begin{equation}
 A_{t_f,{\cal F}_{t_0}}=A({\cal F}_{t_0}\cup \{(x_1,t_1),
 \dots,(x_N,t_N)\})
\end{equation}
This limit is nothing but a Euclidean path integral
\begin{equation}\label{BrPI}
 \mathbb{E}\left( A_{t_f} \right)_{{\bf P},{\cal F}_{t_0}}=
 \int_{x(t_0) = x_0} {\cal D}x~\exp(-S)~A_{t_f,{\cal F}_{t_0}}
\end{equation}
where ${\cal D}x$ includes a properly normalized measure (see below), and
\begin{equation}
 S[x] = \int_{t_0}^{t_f} {{\dot x}^2(t)\over 2}~dt
\end{equation}
is the Euclidean action functional
for a free particle on ${\bf R}$ (as before, ${\dot x}(t) = dx(t)/dt$).

{}Let us note some straightforward differences in units between the path integral in (\ref{BrPI}) and the quantum mechanical Euclidean path integral (\ref{EPI}). Here we have no mass $m$ or $\hbar$, and $x(t)$ does not have the dimension of length but of $\sqrt{t}$. I.e., the two path integrals are the same in the units where $m=\hbar=1$. In these units the discretized measures in (\ref{discrete.E}) and (\ref{discrete.Br}) are identical. Note that the measure in (\ref{discrete.Br}) is a corollary of the measure ${\bf P}$ for the Brownian motion.\footnote{\, Recall that the measure (\ref{discrete.E}) is fixed by requiring agreement with experiment, be it directly or via its derivation using, e.g., Schr\"odinger's equation, which itself is verified experimentally.}

{}In the context of quantum mechanics we considered paths where both endpoints are fixed. Here too we have the following conditional expectation:
\begin{eqnarray}
 \mathbb{E}\left( A_{t_f} \right)_{{\bf P},{\cal F}_{t_0},x(t_f) = x_f} &=&
 \lim_{N\to\infty}~ \prod_{i=1}^{N-1} \int_{-\infty}^\infty dx_i \prod_{i=1}^N {1\over
 \sqrt{2\pi \Delta t_i}}\exp\left(-{(\Delta x_i)^2\over 2\Delta t_i}\right)~
 A_{t_f,{\cal F}_{t_0}} \nonumber\\
 &=&\int_{x(t_0) = x_0,~x(t_f) = x_f} {\cal D}x~\exp(-S)~A_{t_f,{\cal F}_{t_0}}\label{discrete.Br.1}
\end{eqnarray}
Note that
\begin{equation}
 \mathbb{E}\left( A_{t_f} \right)_{{\bf P},{\cal F}_{t_0}} = \int_{-\infty}^\infty dx^\prime~\langle A_{t_f} \rangle_{{\bf P},{\cal F}_{t_0},x(t_f) = x^\prime}
\end{equation}
In the asset pricing context we are mostly interested in this conditional expectation.

{}For $A_t \equiv 1$ the conditional expectation (\ref{discrete.Br.1}) is nothing but the probability\footnote{\, A.k.a. transition density, which is real-valued. Note that in the quantum mechanical context this quantity has the interpretation of the probability {\em amplitude} instead, and it is complex-valued.} of starting at $(x_0,t_0)$ and ending at $(x_f,t_f)$:
\begin{equation}\label{gauss}
 \mathbb{E}\left( 1 \right)_{{\bf P},{\cal F}_{t_0},x(t_f) = x_f}  = P(x_0,t_0; x_f, t_f) = {1\over\sqrt{2\pi(t_f-t_0)}}\exp\left(-{(x_f-x_0)^2\over 2(t_f-t_0)}\right)
\end{equation}
So, Euclidean path integral with $m=\hbar=1$ describes Brownian motion with unit volatility. And we can use the (Euclidean) path integral machinery in asset pricing.

\section{Relativistic Theory}\label{sec.4}

{}From the get-go in Section \ref{sec.2} we assumed non-relativistic dynamics for our classical particle. Thus, the kinetic energy is given by the non-relativistic expression (\ref{kinetic}), force $F$ acts on our particle instantaneously, etc. We then saw that Euclidean quantum mechanics of such a particle describes Brownian motion, which appears in asset pricing, with its (widely accepted and used) Gaussian probability density (\ref{gauss}).

{}Naturally, we can ask: what if we consider a relativistic particle instead? What kind of a probability density will we get? The answer turns out to be intriguing...

\subsection{Action for Relativistic Particle}

{}For our purposes here it will suffice to focus on a free massive relativistic particle, with no forces acting on it. Its action is given by
\begin{equation}\label{rel.ac}
 S = \int dt~L
\end{equation}
where the Lagrangian
\begin{equation}\label{rel.lag}
 L = -mc^2\sqrt{1 - {v^2\over c^2}}
\end{equation}
Here, as before, $v = \dot{x}$ is the velocity, $m$ is the mass, and $c$ is a constant with the dimensionality of velocity. In the real world, $c$ is the speed of light. We will come back to the (geometric) interpretation of the action (\ref{rel.ac}) below. Thus, the reader may recognize the square root in (\ref{rel.lag}) as the Lorentz contraction factor. For now, we simply take the relativistic action (\ref{rel.ac}) as a given and see what we get out of it.

{}First, at velocities well below the speed of light, i.e., when $v^2 \ll c^2$, we can approximate the Lagrangian via $L \approx - mc^2 + {1\over 2}mv^2$. The constant term does not affect the Euler-Lagrange equations (\ref{EL}). However, the second term is precisely the Lagrangian (\ref{nr.lag}) for a non-relativistic particle (with $V=0$ and no force acting on it).

\subsection{Euclidean Relativistic Particle}

{}Second, in (\ref{rel.lag}) the velocity $v$ cannot exceed the speed of light as otherwise the Lagrangian becomes imaginary. However, this restriction does not apply in Euclidean time. Recall that for our purposes in the context of asset pricing, we must Wick-rotate to Euclidean time. Then our action
\begin{equation}\label{rel.v.ac}
 S = \int dt~L = mc^2 \int dt~\sqrt{1 + {v^2\over c^2}}
\end{equation}
There is no restriction on the velocity $v$, it can exceed $c$, which now has an interpretation of a relative normalization between the time coordinate $t$ and the space coordinate $x$. In fact, if we define $y = ct$, then we can write our action as
\begin{equation}
 S = \int dt~L = mc \int d\rho
\end{equation}
where
\begin{equation}
 d\rho^2 = dx^2 + dy^2
\end{equation}
I.e., the spatial coordinate $x$ and time coordinate $y$ are on a completely equal footing and the system, which lives in the 2-dimensional Euclidean space ${\bf R}^2$, possesses a 2-dimensional $SO(2)$ rotational invariance.\footnote{\, If we had $d$ spatial coordinates $x_i$, $i=1,\dots,d$, instead of just one, we would end up with $(d+1)$-dimensional rotational invariance (under the group $SO(d+1)$) in the $(d+1)$-dimensional Euclidean space ${\bf R}^{d+1}$. See, e.g., \cite{Kleinert} for a detailed discussion.} Space and time are indistinguishable.

\subsection{Reparametrization Invariance}

{}In fact, our action has a local reparametrization symmetry. Let us denote $x^\mu(\tau) = (x(\tau), y(\tau))$, where $\tau$ parameterizes the ``worldline" $x^\mu(\tau)$. Then we have (summation over repeated indices is implicit)
\begin{eqnarray}\label{rep.ac}
 &&S = mc\int d\tau~\sqrt{x^{\prime 2}(\tau)}\\
 &&x^{\prime 2}(\tau) = \delta_{\mu\nu}{dx^\mu(\tau)\over d\tau}{dx^\nu(\tau)\over d\tau}
\end{eqnarray}
The $\mu,\nu = 1,2$ indices are contracted with the Euclidean metric $\delta_{\mu\nu}$ (Kronecker delta). The action (\ref{rep.ac}) is invariant under local reparametrizations
\begin{equation}\label{rep}
 \tau \rightarrow f(\tau)
\end{equation}
where $f(\tau)$ is an arbitrary (differentiable) function.

{}One issue with the action (\ref{rep.ac}) is that it is highly nonlinear in $x^{\prime 2}$. There is a neat trick to circumvent this. We can introduce an auxiliary Lagrange multiplier (a.k.a. an ``einbein") $e(\tau)$, an extra degree of freedom on the particle worldline, and rewrite the action (\ref{rep.ac}) via
\begin{equation}\label{e.ac}
 S = \int d\tau \left[{1\over e(\tau)}{mx^{\prime 2}(\tau)\over 2} + e(\tau){mc^2\over 2}\right]
\end{equation}
The equation of motion for $e(\tau)$ gives
\begin{equation}
 e(\tau) = {1\over c} \sqrt{x^{\prime 2}(\tau)}
\end{equation}
Therefore, integrating out $e(\tau)$, we arrive at (\ref{rep.ac}). Under the reparametrizations (\ref{rep}) we have
\begin{equation}
 e(\tau) \rightarrow {e(\tau)\over f^\prime(\tau)}
\end{equation}
Therefore, the only reparametrization-invariant quantity we can construct to parameterize the worldline is its length ($\tau_0$ and $\tau_f$ are the initial and final values of $\tau$ on the worldline):
\begin{equation}
 s = \int_{\tau_0}^{\tau_f} d\tau~e(\tau)
\end{equation}
This implies that we can completely gauge-fix the action (\ref{e.ac}), i.e., we can completely fix the reparametrization symmetry by setting $e(\tau) \equiv 1$, so $s = \tau_f - \tau_0$. The gauge-fixed action reads:
\begin{equation}\label{gauge.ac}
 S_{fixed} = \int_0^s d\tau \left[{mx^{\prime 2}(\tau)\over 2} + {mc^2\over 2}\right]
\end{equation}
This is nothing but the action for a {\em two}-dimensional {\em non-relativistic} Euclidean free particle with a constant potential.\footnote{\, The one-dimensional non-relativistic Euclidean Lagrangian with potential $V$ is given by (\ref{1D.lag}).} Now we can readily compute the path integral.

\subsection{Path Integral}

{}As in the non-relativistic case, here we are after the transition density, i.e., the probability of starting at $(x_0,t_0)$ and ending at $(x_f, t_f)$, except now we have a relativistic Euclidean particle. This probability, call it ${\widetilde P}(x_0,t_0;x_f,t_f)$, is given by the path integral (we keep $m$ and $\hbar$ arbitrary for now, we will set them to 1 below -- we are keeping them around for a reason)
\begin{eqnarray}
 &&{\widetilde P}(x_0,t_0;x_f,t_f) = {\cal N}\int_0^\infty ds\times \nonumber\\
 &&\,\,\,\,\,\,\,\times \int_{x^1(\tau_0) = x_0,~x^2(\tau_0) = ct_0,~x^1(\tau_f) = x_f,~x^2(\tau_f) = ct_f}{\cal D}x^1{\cal D}x^2 \exp(-{S_{fixed}\over \hbar})
\end{eqnarray}
Here: ${\cal N}$ is an overall normalization factor which we will fix below; the path integrals over the 2 coordinates ${\cal D}x^\mu$, happily, are Gaussian and independent of each other; we must also integrate over all possible path lengths, hence the integral over $ds$.

{}The ${\cal D}x^1$ and ${\cal D}x^2$ integrals simply produce Gaussian transition densities as in (\ref{gauss}),\footnote{\, Except in (\ref{gauss}) $t$ is now replaced by $\tau$ and $x$ by $x^1$ or $x^2$, accordingly, and $m$ and $\hbar$ are restored.} so we get
\begin{equation}\label{tr.den.s}
 {\widetilde P}(x_0,t_0;x_f,t_f) = {\cal N}\int_0^\infty ds~{m\over 2\pi\hbar s}~\exp\left(-{mz^2\over 2\hbar s} - {mc^2s\over 2\hbar}\right)
\end{equation}
where $z^2 = \delta_{\mu\nu}z^\mu z^\nu = (x_f-x_0)^2 + c^2(t_f - t_0)^2$. Let us Fourier transform the $z^\mu$ dependence:
\begin{equation}
 {\widetilde P}(x_0,t_0;x_f,t_f) = {\cal N} \int_0^\infty ds~\exp\left(- {mc^2s\over 2\hbar}\right)\int {d^2k\over (2\pi)^2}
 ~\exp\left(ik_\mu z^\mu -{k^2\hbar s\over 2m}\right)
\end{equation}
where $d^2 k = dk_1dk_2$ and $k^2 = \delta^{\mu\nu}k_\mu k_\nu$. Integrating over $s$ we get
\begin{equation}\label{mom.rep}
 {\widetilde P}(x_0,t_0;x_f,t_f) = {2{\cal N}m\over\hbar} \int {d^2k\over (2\pi)^2}~{\exp\left(ik_\mu z^\mu\right)\over {k^2 + m^2 c^2/\hbar^2}}
\end{equation}
Therefore, (for now -- see below) setting ${\cal N} = \hbar/2m$, ${\widetilde P}(x_0,t_0;x_f,t_f) = {\widetilde P}(z^\mu)$ is nothing but the 2-dimensional Euclidean Green's function for the Klein-Gordon equation\footnote{\, Since the spacetime is Euclidean, it is actually a Green's function for the 2D Laplacian.}
\begin{equation}\label{KG}
 \left(-\Box  + m^2 c^2/\hbar^2\right) {\widetilde P}(z^\mu) = \delta^{(2)}(z^\mu)
\end{equation}
where $\Box = \delta^{\mu\nu}\partial_\mu\partial_\nu$ (while $\partial_\mu = \partial/\partial z^\mu$, and $\delta^{(2)}(z^\mu) = \delta(z^1)\delta(z^2)$). Therefore, we have\footnote{\, The $D$-dimensional Euclidean Green's function for the Klein-Gordon equation $G^{(D)}(z^\mu) = {1\over (2\pi)^{D/2}}~\left({mc\over \hbar\sqrt{z^2}}\right)^{D/2-1}K_{D/2-1}\left({mc\sqrt{z^2}\over\hbar}\right)$. See, e.g., \cite{Kleinert} for a detailed discussion.}
\begin{equation}
 {\widetilde P}(x_0,t_0;x_f,t_f) = {1\over 2\pi} K_0\left({mc\over\hbar}\sqrt{(x_f-x_0)^2 + c^2(t_f-t_0)^2}\right)
\end{equation}
With the normalization ${\cal N} = \hbar/2m$, this result follows directly from (\ref{tr.den.s}) via the substitution $\xi = \ln\left(cs/\sqrt{z^2}\right)$ and recalling the integral representation
\begin{equation}\label{K_alpha}
 K_\alpha(x) = \int_0^\infty d\xi~\exp\left(-x\cosh(\xi)\right)\cosh(\alpha\xi)
\end{equation}
Note that ${\widetilde P}(x_0,t_0;x_f,t_f) = {\widetilde P}(z^\mu)$ is invariant under 2-dimensional $SO(2)$ rotations. However, as we will see below, this invariance is broken in the financial context.

\subsection{Normalization and Effect of Rest Energy}

{}Above, ad hoc, we set the normalization factor ${\cal N} = \hbar/2m$ just so that ${\widetilde P}(z^\mu)$ coincides with the 2-dimensional Euclidean Green's function. However, in the financial context what we care about is that ${\widetilde P}(x_0,t_0;x_f,t_f)$ has the correct interpretation, to wit, as the transition density (once we set $m = \hbar = 1$, that is). In particular, it must obey the normalization condition
\begin{equation}\label{norm}
 \int_{-\infty}^\infty dx_f {\widetilde P}(x_0,t_0;x_f,t_f) \equiv 1
\end{equation}
Note that the non-relativistic transition density (\ref{gauss}) has this property. However, the relativistic ${\widetilde P}(x_0,t_0;x_f,t_f)$, as normalized above, does not, and for a good reason.

{}The simplest way to calculate the integral on the l.h.s. of (\ref{norm}) is via (\ref{mom.rep}) as the integral over $dx_f/2\pi$ produces a delta-function $\delta(k_1)$, which gets rid of the $dk_1$ integral leaving us with (recall that $t_f \geq t_0$)
\begin{eqnarray}
 \int_{-\infty}^\infty dx_f {\widetilde P}(x_0,t_0;x_f,t_f) &=& {2{\cal N}m\over\hbar} \int {dk_2\over 2\pi}~{\exp\left(ik_2c\left[t_f-t_0\right]\right)\over {k_2^2 + m^2c^2/\hbar^2}} =\nonumber\\
 &=&{{\cal N}\over c}~\exp\left(-{mc^2\over\hbar}\left[t_f - t_0\right]\right)\label{rest.mass}
\end{eqnarray}
So, to get the correct normalization, we must set
\begin{equation}\label{cor.norm}
 {\cal N} = c~ \exp\left({mc^2\over\hbar}\left[t_f - t_0\right]\right)
\end{equation}
It will now become clear why we kept $m$ and $\hbar$ around instead of setting $m=\hbar=1$.

{}The exponential decay factor on the r.h.s. of (\ref{rest.mass}) is due to Einstein's famous rest energy $E_0 = mc^2$ of a relativistic particle, which is not included in the non-relativistic case. Indeed, in the non-relativistic limit, when $v^2 \ll c^2$, the action (\ref{rel.v.ac}) reads
\begin{equation}
 S = mc^2(t_f - t_0) + \int_{t_0}^{t_f} dt~{mv^2\over 2}\left[1 + {\cal O}\left({v^2\over c^2}\right)\right]
\end{equation}
We can subtract the rest energy term by considering a modified action
\begin{equation}
 S_{mod} = mc^2\int dt \left[\sqrt{1 + {v^2\over c^2}} - 1\right]
\end{equation}
However, this action, unlike (\ref{rel.v.ac}), cannot be made reparametrization invariant via the above procedure, which thus no longer applies. We will come back to this below.

{}For now, we will simply set ${\cal N}$ via (\ref{cor.norm}). Then we have\footnote{\, This is an example of ``generalized hyperbolic distributions" \cite{Eberlein}.}
\begin{equation}
 {\widetilde P}(x_0,t_0;x_f,t_f) = {mc\over \pi\hbar} \exp\left({mc^2\over\hbar}\left[t_f - t_0\right]\right) K_0\left({mc\over\hbar}\sqrt{(x_f-x_0)^2 + c^2(t_f-t_0)^2}\right)
\end{equation}
Using the asymptotic property $K_\alpha(x) \sim \sqrt{\pi/2x}\exp(-x)$ for $x\gg 1$, in the non-relativistic limit $c\rightarrow \infty$ we obtain:
\begin{equation}\label{non-rel}
 {\widetilde P}(x_0,t_0;x_f,t_f) \approx \sqrt{m\over 2\pi\hbar\left(t_f-t_0\right)} \exp\left(-{m\left(x_f-x_0\right)^2\over 2\hbar\left(t_f-t_0\right)}\right)
\end{equation}
which is nothing but the Gaussian transition density (and we can set $m=\hbar=1$).

\subsection{Financial Interpretation}

{}From (\ref{non-rel}) it is evident that $\hbar/m$ in the financial context is replaced by $\sigma^2$, where $\sigma$ is the diffusion coefficient (volatility) of the Brownian motion in the non-relativistic limit. We will now interpret the speed of light $c$ and the exponential decay factor on the r.h.s. of (\ref{rest.mass}). Let us for now set the normalization factor ${\cal N} = c$, so we have
\begin{equation}\label{P_K0}
 {\widetilde P}(x_0,t_0;x_f,t_f) = {c\over \pi\sigma^2} K_0\left({c\over\sigma^2}\sqrt{(x_f-x_0)^2 + c^2(t_f-t_0)^2}\right)
\end{equation}
and
\begin{equation}\label{norm.r0}
 \int_{-\infty}^\infty dx_f {\widetilde P}(x_0,t_0;x_f,t_f) = \exp\left(-r_0\left[t_f - t_0\right]\right)
\end{equation}
where
\begin{equation}
 r_0 = {c^2\over \sigma^2}
\end{equation}
The system is invariant under the simultaneous rescalings $x\rightarrow \lambda x$, $c \rightarrow \lambda c$, $\sigma\rightarrow \lambda \sigma$ and ${\widetilde P}\rightarrow {\widetilde P}/\lambda$ (and $r_0$ is invariant). If we take $x$ to be dimensionless, then $c$ has dimension of $1/t$, and $\sigma$ has dimension of $1/\sqrt{t}$. Then $T_* = 1/\sigma^2$ and $v_* = \sigma^2$ define the characteristic time and ``speed" of the diffusion, respectively. If $\sigma^2 \ll c$, then relativistic effects are small and can be ignored. However, if $\sigma^2 \gg c$, then we are in the ``ultra-relativistic" regime. Generally, relativistic effects cannot be ignored when $\sigma^2\gsim c$. So, $c$ is the diffusion ``speed" delineating the non-/relativistic regimes.

{}What about $r_0$? It is simply a constant {\em interest rate} which is always present in the relativistic case, as is the rest energy $E_0 = mc^2$ in Special Relativity. So, we have {\em un problema peque\~{n}ito}. Thus, suppose our relevant time horizon is $T$. If we assume that the interest rate $r_0$ is small, i.e., $r_0 T \ll 1$, then we have $c/\sigma^2 = \sqrt{r_0 T}/\sqrt{\sigma^2 T} \ll 1/\sqrt{\sigma^2 T}$. For the volatility effects to be sizable, we must have $\sigma^2 T\gsim 1$, and in this case we invariably have $c \ll \sigma^2$, i.e., we are in the ``ultra-relativistic regime". This is not necessarily all that bad -- the non-relativistic regime is not interesting as we already know how to deal with it. However, the borderline relativistic regime $c \sim \sigma^2$ is unavailable unless $r_0 T\sim 1$ or we somehow eliminate this built-in interest rate $r_0$.

{}However, a far more serious issue is that the differential operator on the l.h.s. of (\ref{KG}) is second-order w.r.t. time derivatives. It is therefore unclear how to construct martingales under the measure ${\widetilde P}(z^\mu)$ -- and this is irrespective of the nuisance caused by the presence of $r_0$. Happily, the measure ${\widetilde P}(z^\mu)$ does satisfy a Fokker-Planck-like equation, which is first-order w.r.t. time derivatives. In fact, understanding how to arrive at this equation, among other things, will also help get rid of unappealing $r_0$.

\subsection{Hamiltonian Approach}

{}Away from the origin, we can rewrite (\ref{KG}) (we have self-explanatorily simplified the notations)
\begin{equation}
 \partial_t^2 {\widetilde P}(x, t)  = \left(-c^2\partial_x^2 + m^2c^4/\hbar^2\right){\widetilde P}(x, t)
\end{equation}
via a first-order (in time derivatives) equation
\begin{equation}\label{FP}
 \hbar~\partial_t {\widetilde P}(x, t)  = - {\widehat H}~{\widetilde P}(x, t)
\end{equation}
Here the Hamiltonian
\begin{equation}\label{Ham}
 {\widehat H} = \sqrt{-c^2\hbar^2\partial_x^2 + m^2c^4}
\end{equation}
is a nonlocal (w.r.t. $x$) operator, which is understood to act on a given function
\begin{equation}
 f(x, t) = \int {dk\over 2\pi\hbar}~\exp(ikx/\hbar)~{\widetilde f}(k, t)
\end{equation}
via (${\widetilde f}(k, t)$ is the Fourier transform of $f(x, t)$) \cite{Newton}:
\begin{equation}
 {\widehat H}f(x, t) = \int {dk\over 2\pi\hbar}~\exp(ikx/\hbar)~\sqrt{c^2k^2 + m^2c^4}~{\widetilde f}(k, t)
\end{equation}
Alternatively, we can think of ${\widehat H}$ as an infinite power series of differential operators
\begin{equation}
 {\widehat H} = \sqrt{m^2c^4 - c^2{\widehat p}^2} = mc^2\left[1 - {1\over 2}{{\widehat p}^2\over m^2 c^2} - {1\over 8}{{\widehat p}^4\over m^4 c^4} + \dots\right]
\end{equation}
where ${\widehat p} = - \hbar\partial_x$ is the momentum operator (in Euclidean time). Also, note that, with the sign on the r.h.s. of (\ref{FP}) fixed, we have a choice of the sign on the r.h.s. of (\ref{Ham}). However, we must choose the $+$ sign in (\ref{Ham}), or else we will get incorrect asymptotic behavior at infinity (that is, $I_\alpha(\cdot)$ instead of $K_\alpha(\cdot)$ -- see below).

{}To see that ${\widetilde P}(x, t)$ satisfies (\ref{FP}), we can integrate over $dk_2$ in (\ref{mom.rep}) and obtain (here we have a single integral over $dk$, and also $t > 0$)
\begin{equation}\label{P.rel}
 {\widetilde P}(x, t) = {{\cal N}m\over\hbar} \int {dk\over 2\pi} \exp\left(ikx\right) {\exp\left(-ct\sqrt{k^2 + m^2c^2/\hbar^2}\right)\over\sqrt{k^2 + m^2c^2/\hbar^2}}
\end{equation}
In this representation it is evident that ${\widetilde P}(x, t)$ satisfies (\ref{FP}). Also, this fixes the precise form of the function $\Omega(k)$ in the general solution
\begin{equation}\label{P.Ham}
 {\widetilde P}(x, t) = \exp\left(-{t\over \hbar}{\widehat H}\right) \int {dk\over 2\pi} \exp\left(ikx\right) \Omega(k)
\end{equation}
to (\ref{FP}), to wit,
\begin{equation}\label{Omega}
 \Omega(k) = {{\cal N}m\over\hbar}~{1\over\sqrt{k^2 + m^2c^2/\hbar^2}}
\end{equation}
thanks to our computation of ${\widetilde P}(x, t)$ using path integral above. However, we still have aforesaid {\em un problema peque\~{n}ito}. If we set ${\cal N} = c$, we get (\ref{norm.r0}) due to the rest energy contribution to ${\widehat H}$. Except that now we are in good shape to fix this.

\subsubsection{Removing Rest Energy}

{}To remove the rest energy contribution, we can simply subtract $E_0 = mc^2$ from the Hamiltonian. I.e., we still have (\ref{P.Ham}), except that the Hamiltonian is now given by\footnote{\, This Hamiltonian, in the lingo of L{\'e}vy processes, is discussed in, e.g., \cite{Applebaum}.}
\begin{equation}\label{Ham.fixed}
 {\widehat H} = \sqrt{m^2c^4 - c^2{\widehat p}^2} - mc^2 = mc^2\left[ - {1\over 2}{{\widehat p}^2\over m^2 c^2} - {1\over 8}{{\widehat p}^4\over m^4 c^4} + \dots\right]
\end{equation}
In the non-relativistic limit ($c\rightarrow\infty$) we now have the usual expression\footnote{\, The ``extra" minus sign is no reason for alarm: we are in Euclidean time.} ${\widehat H} \approx -{\widehat p}^2/2m$ without the rest energy contribution. However, the assumption that $\Omega(k)$ is given by (\ref{Omega}) is no longer justified. Indeed, we derived (\ref{Omega}) using the path integral approach, where we started with a reparametrization invariant action, fixed the gauge, etc. Once we modify the Hamiltonian as in (\ref{Ham.fixed}), the reparametrization invariance is gone (and so is the $SO(2)$ rotational invariance in the $(x,y)$ plane, where $y=ct$), so that entire derivation is no longer applicable. But not all is lost.

\subsubsection{Tower Law}

{}We can start directly from the Fokker-Planck equation (\ref{FP}) with ${\widehat H}$ given by (\ref{Ham.fixed}). The general solution is given by (\ref{P.Ham}). However, a priori $\Omega(k)$ is arbitrary apart from the usual requirement that ${\widetilde P}(x, t)$ properly decay at infinity. Nonetheless, we can fix $\Omega(k)$ by requiring that ${\widetilde P}(x,t)$ obey the {\em tower law} for conditional expectations.\footnote{\, A.k.a. the law of iterated expectation. See \cite{Samuelson1}, \cite{Samuelson2}.}

{}Generally, the tower law (see, e.g., \cite{Baxter})\footnote{\, For detailed lecture notes, see, e.g., \cite{Phynance}.} states that
\begin{equation}
 \mathbb{E}\left(\mathbb{E}\left( X_T \right)_{{\bf P},{\cal F}_t}\right)_{{\bf P},{\cal F}_s} = \mathbb{E}\left( X_T \right)_{{\bf P},{\cal F}_s},~~~0\leq s\leq t\leq T
\end{equation}
where ${\bf P}$ is some measure. The tower law implies that the process
\begin{equation}
 N_t = \mathbb{E}\left( X_T \right)_{{\bf P},{\cal F}_t}
\end{equation}
is a ${\bf P}$-martingale (provided $\mathbb{E}\left(|X_T|\right)_{\bf P}$ is finite). By definition, a process $M_t$ (with finite $\mathbb{E}\left(|M_t|\right)_{\bf P}$) is a ${\bf P}$-martingale if and only if for all $t\geq 0$
\begin{equation}
 \mathbb{E}\left( M_t \right)_{{\bf P},{\cal F}_s} = M_s,~~~0\leq s\leq t
\end{equation}
The property that conditional expectations are martingales -- which hinges on the tower law -- is paramount in option (and generally contingent claim) pricing, to wit, in constructing a self-financing hedging strategy for a claim $X_T$ at maturity $T$.

{}Therefore, we require that our measure ${\widetilde P}(x, t)$ obey the tower law. Note that originally we wrote this measure as ${\widetilde P}(x_0,t_0;x_f,t_f)$ and then simplified notations. Because of the homogeneity (i.e., translational invariance) in both $x$ and $t$ directions, we have
${\widetilde P}(x_0,t_0;x_f,t_f) = {\widetilde P}(x_f - x_0, t_f - t_0)$, hence the condensed notations. The tower law then translates to
\begin{equation}\label{tower}
 \int_{-\infty}^\infty dx^{\prime\prime}~{\widetilde P}\left(x - x^{\prime\prime}, t - t^{\prime\prime}\right) {\widetilde P}\left(x^{\prime\prime} - x^\prime, t^{\prime\prime}- t^\prime\right) = {\widetilde P}\left(x - x^\prime, t - t^\prime\right)
\end{equation}
where $t^{\prime\prime}$ is arbitrary so long as $t^\prime \leq t^{\prime\prime} \leq t$. We will now derive a condition on $\Omega(k)$ from (\ref{tower}) for a general class of Hamiltonians ${\widehat H}$. For our purposes here it will suffice to consider Hamiltonians of the form ${\widehat H} = {\cal H}(-\partial^2_x)$, i.e., we only have even derivatives.\footnote{\, So, for our Hamiltonian (\ref{Ham.fixed}) we have ${\cal H}(u) = \sqrt{m^2c^4 + c^2\hbar^2 u} - mc^2$.}  Then plugging (\ref{P.Ham}) into (\ref{tower}) and noting that ${\cal H}(-\partial^2_{x^\prime})f(x - x^\prime) = {\cal H}(-\partial^2_x)f(x - x^\prime)$ ($f(x)$ is some function), we get that the l.h.s. of (\ref{tower}) equals
\begin{equation}
 \int {dk\over 2\pi} \exp\left(ik(x-x^\prime)\right) \Omega^2(k)
\end{equation}
Therefore, the tower law invariably implies that $\Omega(k)\equiv 1$. So, what does this mean?

{}Note that for the non-relativistic measure (\ref{gauss}) we have ${\cal H}(u) = \hbar^2 u/2m$ and $\Omega(k)\equiv 1$, so the tower law is satisfied. However, the ``na\"{i}ve" relativistic measure (which includes the rest energy contribution) we derived above has $\Omega(k)$ given by (\ref{Omega}), so it does not satisfy the tower law and would be unsuitable for option pricing (regardless of the rest energy contribution). This result might appear bizarre; however, in physics it has been known for quite some time that the relativistic propagator (\ref{mom.rep}) does not possess the (tower law) semigroup property (\ref{tower});\footnote{\, More precisely, in physics this propagator usually is considered in Minkowski spacetime.} see, e.g., \cite{Hartle}, \cite{Redmount}, and references therein.\footnote{\, This issue can be traced to the definition of the path integral measure. The reparametrization and $SO(2)$ rotational invariance require the measure such that we have (\ref{mom.rep}).} In physics this is interpreted as inadequacy of the so-called first-quantized theory (i.e., relativistic quantum mechanics) and necessity of the so-called second-quantized theory (i.e., quantum field theory) for describing relativistic particles. In the context of finance such dramatic measures might appear as overkill.\footnote{\, However, it is tricker than that. We will come back to this below.} We can simply choose to work with the measure with $\Omega(k)\equiv 1$. The only ``issue" with this choice is that the $SO(2)$ rotational invariance broken, i.e., space and time coordinates are no longer indistinguishable. However, in the context of finance this is a non-issue and in fact a welcome feature: empirically there is no evidence of such $SO(2)$ invariance, so it had better be broken! Thus, as little as introducing an interest rate into the system would invariably break such invariance. So, our pragmatic approach here is to take the probability density ${\widetilde P}(x, t)$ corresponding to $\Omega(k)\equiv 1$ and understand the effects and implications of the higher derivative terms in the Hamiltonian (\ref{Ham.fixed}) in the context of option pricing, etc. I.e., our probability measure is given by
\begin{equation}\label{P.corr}
 {\widetilde P}(x, t) = \int {dk\over 2\pi} \exp\left(ikx\right)
 \exp\left(-ct\left[\sqrt{k^2 + m^2c^2/\hbar^2} - mc/\hbar\right]\right)
\end{equation}
which can be evaluated by noting that (\ref{P.rel}) (with ${\cal N} = c$) is the same as (\ref{P_K0}) (with $\sigma^2 = \hbar/m$), that $K_1(x) = -K^\prime_0(x)$ (which follows from (\ref{K_alpha})), and that (\ref{P.corr}) up to an overall factor of $-(mc^2/\hbar)\exp(mc^2t/\hbar)$ equals the first time derivative of (\ref{P.rel}), so we get for (\ref{P.corr})
\begin{equation}\label{P_K1}
 {\widetilde P}(x, t) = {mc\over\pi \hbar} {ct\over\sqrt{x^2 + c^2t^2}} K_1\left({mc\over\hbar}\sqrt{x^2 + c^2 t^2}\right)\exp(mc^2 t/\hbar)
\end{equation}
Apart from the last factor of $\exp(mc^2t/\hbar)$ (which removes the rest energy contribution), this is the Euclidean version of the Newton-Wigner propagator first derived in \cite{Newton} (also see \cite{Hartle}).\footnote{\, This is an example of ``generalized hyperbolic distributions" \cite{Eberlein} referred to as the normal inverse Gaussian distribution \cite{BN}, \cite{CarrNIG}, \cite{Drag}. I am grateful to Jean-Philippe Bouchaud for the last reference.} Our probability density ${\widetilde P}(x, t)$ given by (\ref{P_K1}) has the property that in the non-relativistic limit ($c\rightarrow \infty$) we still recover the Gaussian probability density (\ref{non-rel}).\footnote{\, This is because the leading large $x\gg 1$ behavior of $K_\alpha(x)$ is $K_\alpha(x) \sim \sqrt{\pi/2x}\exp(-x)$ irrespective of $\alpha$. So, we can readily construct an infinite number of probability densities with the correct non-relativistic limit by taking higher time derivatives of (\ref{P_K0}). However, the tower law plus the normalization requirement (see below) uniquely fix the probability density to be given by (\ref{P_K1}).\label{fn.K}} Furthermore, we have the correct normalization property
\begin{equation}
 \int_{-\infty}^\infty dx~{\widetilde P}(x, t) = 1
\end{equation}
It then follows that the identity process $I_t\equiv 1$ is a (trivial) martingale. Finally, in the financial context, our probability density reads
\begin{equation}
 {\widetilde P}(x, t) = {c\over\pi\sigma^2} {ct\over\sqrt{x^2 + c^2t^2}} K_1\left({c\over\sigma^2}\sqrt{x^2 + c^2 t^2}\right)\exp(c^2 t/\sigma^2)
\end{equation}
where we have identified $\hbar/m$ with $\sigma^2$, $\sigma$ being the diffusion coefficient (volatility).

\subsection{Martingales}

{}We must now construct nontrivial martingales that correspond to (discounted) stock prices. For the sake of simplicity, we will set the risk-free interest rate to zero -- we will straightforwardly restore it at the end. The process $S_t = S_0\exp(\sigma x(t) - \sigma^2t/2)$ is a martingale under the Gaussian measure (\ref{gauss}), but not under our non-Gaussian measure ${\widetilde P}(x, t)$ given by (\ref{P_K1}). However, in the Hamiltonian formalism it is straightforward to (at least {\em formally} -- see below) construct nontrivial martingales even if the measure is non-Gaussian \cite{Kleinert2}. By definition, the process $S_t = S(x(t), t)$ is a martingale if and only if (assuming the expected value of $|S_{t^\prime}|$ is finite)
\begin{equation}
 \int_{-\infty}^\infty dx^\prime~{\widetilde P}(x^\prime - x, t^\prime - t)~S(x^\prime, t^\prime) = S(x, t)
\end{equation}
where $t^\prime \geq t$. As above, let us assume that ${\widehat H} = {\cal H}(-\partial_x^2)$. By virtue of (\ref{FP}) we then have the following differential equation for $S(x, t)$:
\begin{equation}
 \hbar~\partial_t S(x, t) = {\widehat H}~S(x, t)
\end{equation}
A nontrivial solution is given by
\begin{equation}
 S(x, t) = S_0\exp\left(\zeta x + {\cal H}(-\zeta^2)t/\hbar\right)
\end{equation}
where $\zeta$ is a priori arbitrary. In the non-relativistic case we have ${\cal H}(u) = \hbar^2 u / 2m$, so we get the familiar martingale
\begin{equation}\label{nonrel.mart}
 S(x, t) = S_0\exp\left(\zeta x -\zeta^2\sigma^2 t/2\right)
\end{equation}
where $\sigma^2 = \hbar/m$. For our Hamiltonian (\ref{Ham.fixed}) we have ${\cal H}(u) = \
\sqrt{m^2c^4 + c^2\hbar^2u} - mc^2$, so our martingale reads:
\begin{equation}\label{rel.mart}
 S(x, t) = S_0 \exp\left(\zeta x - {c^2 t \over \sigma^2} \left[1 - \sqrt{1 - {\zeta^2 \sigma^4 \over c^2}}\right]\right)
\end{equation}
which in the $c\rightarrow \infty$ limit reduces to the non-relativistic martingale given by (\ref{nonrel.mart}).

{}A peculiar feature of the martingale (\ref{rel.mart}) is that there is a restriction on the log-volatility $\sigma_S = \zeta\sigma$. Thus, as above, let $x(t)$ be dimensionless, then so is $\zeta$. From (\ref{rel.mart}) we have
\begin{equation}\label{max.vol}
 \sigma^2_S \leq c^2/\sigma^2
\end{equation}
so if $\sigma^2 > c$, then invariably $\sigma_S^2 < c$. To get $\sigma_S^2 > c$, we must have $\sigma^2 < c$. However, for given $c$ and $\sigma$ we have a bound (\ref{max.vol}). This peculiarity is absent in the non-relativistic case as there $\sigma$ and $\zeta$ always appear in the combination $\sigma_S = \zeta\sigma$. This is not so in the relativistic case as we have another dimensionful quantity $c$.\footnote{\, Mathematically, the bound (\ref{max.vol}) stems from the asymptotic behavior at large $|x|$ (see fn. \ref{fn.K}): ${\widetilde P}(x,t)$ decays as $\exp(-c|x|/\sigma^2)/x^{3\over 2}$ (cf. the Gaussian density), hence the requirement $|\zeta|\leq c/\sigma^2$.}

\subsubsection{Tachyonic Martingales}

{}So, if the diffusion is ``superluminal" or tachyonic (borrowing physics terminology), i.e., $\sigma^2 > c$, our martingales (\ref{rel.mart}) have ``subluminal" log-volatility $\sigma_S^2 < c$. Can we construct other -- tachyonic -- martingales in this case without such a restriction? The answer is yes, but it is tricky. Thus,
\begin{equation}\label{tach.mart}
 S(x, t) = S_0 \left[a\cos\left(\zeta x\right) + b\sin\left(\zeta x\right)\right]\exp\left({c^2 t \over \sigma^2} \left[\sqrt{1 + {\zeta^2 \sigma^4 \over c^2}} - 1\right]\right)
\end{equation}
is a martingale ($a$ and $b$ are constants). However, this martingale per se is not positive-definite and is unsuitable to describe stock prices. Adding a constant (or a martingale of the form (\ref{rel.mart})) to it does not help as the time-dependent exponent in (\ref{tach.mart}) grows with time. One way we can attempt to circumvent the non-positivity of (\ref{tach.mart}) is to introduce {\em reflecting} boundaries and keep the process positive that way.\footnote{\, For recent discussions and references, see \cite{HoLee}, \cite{FX}.} However, this is a different ball game (if attainable at all) we will not pursue here.

\subsection{Option Pricing, Fat Tails and Volatility Smiles}

{}Instead, we will discuss option pricing for the martingales of the form (\ref{rel.mart}). We are still setting the risk-free interest rate to zero. Let us consider a European call option with the claim $X_T = \left(S_T - S_*\right)^+ = \mbox{max}\left(S_T - S_*, 0\right)$ at maturity $T$, where we use $S_*$ for the strike price (instead of more conventional $k$) not to confuse it with Fourier integration over $dk$. The option price at time $t\leq T$ is given by
\begin{equation}\label{call}
 V_c(S_t, t) = \int_{-\infty}^\infty dx^\prime {\widetilde P}(x^\prime - x, T-t) \left(S_T - S_*\right)^+
\end{equation}
where
\begin{eqnarray}
 &&S_t = S_0 \exp\left(\zeta x - \beta t \right)\\
 &&S_T = S_0 \exp\left(\zeta x^\prime - \beta T\right)\\
 &&\beta = {c^2 \over \sigma^2} \left[1 - \sqrt{1 - {\zeta^2 \sigma^4 \over c^2}}\right]
\end{eqnarray}
Therefore, we can rewrite (\ref{call}) via (without loss of generality we are assuming $\zeta > 0$)
\begin{equation}\label{call.prc}
 V_c(S_t, t) = S_*\int_{x_*}^\infty dx^\prime {\widetilde P}(x^\prime, T-t)\left(\exp(\zeta[x^\prime - x_*]) - 1\right)
\end{equation}
where
\begin{equation}\label{x_*}
 x_* = {1\over\zeta}\left[\ln\left({S_*\over S_t}\right) + \beta(T-t)\right]
\end{equation}
The r.h.s. of (\ref{call.prc}) can be evaluated numerically. However, for our purposes here it suffices to look at the asymptotic behavior of the distribution ${\widetilde P}(x, t)$ at large $x$, which can be approximated as
\begin{equation}\label{fat}
 {\widetilde P}(x, t) \approx {1\over\sqrt{2\pi\sigma^2 t}} \left[1 + {x^2\over c^2t^2}\right]^{-{3\over 4}} \exp\left(-{c^2t\over\sigma^2}\left[\sqrt{1 + {x^2\over c^2 t^2}} - 1\right]\right)
\end{equation}
If, formally, we take the $c\rightarrow \infty$ limit (non-relativistic approximation), we recover the Gaussian density. However, in the relativistic case this approximation apparently is misleading. For large $x$ the decay is much softer and the distribution (\ref{fat}) has fat tails. It then invariably follows (see, e.g., \cite{Hull}) that option prices using the relativistic distribution automatically have volatility smiles. Looks can be deceiving.

\subsubsection{Nonzero Interest Rate}\label{int.rate}

{}Above we assumed zero interest rates. If we have a nonzero constant interest rate $r$, the option pricing formula (\ref{call.prc}) is modified as follows: i) we have an overall factor of $\exp(-r[T-t])$ in front of the integral; and ii) $\beta$ is replaced by $\beta - r$ in (\ref{x_*}).

\subsubsection{Self-financing Hedging Strategy}

{}Above we mention self-financing hedging strategies. The {\em standard} argument\footnote{\, There is a ``caveat" in the standard argument in the context hereof, hence the emphasis.} goes as follows (see, e.g., \cite{Baxter}). Suppose we have a tradable $S_t$ and a numeraire\footnote{\, Usually, the numeraire is chosen to be a cash bond, but it can be any tradable instrument.} $B_t$. Generally, $B_t$ need not be deterministic. Consider a claim $X_T$ at maturity $T$. To value this claim at time $t\leq T$, we need to construct a self-financing hedging strategy which replicates the claim $X_T$. The hedging strategy amounts to holding a portfolio $(\phi_t,\psi_t)$ consisting of $\phi_t$ units of $S_t$ and $\psi_t$ units of $B_t$, where $\phi_t$ and $\psi_t$ are {\em previsible} processes. The value $V_t$ of this portfolio at time $t$ is given by
\begin{equation}\label{V}
 V_t = \phi_t S_t + \psi_t B_t
\end{equation}
The self-financing property means that the change in the value of the portfolio is solely due the changes in the values of $S_t$ and $B_t$, i.e., there is no cash flowing in or out of the strategy at any time:
\begin{equation}\label{self}
 dV_t = \phi_t dS_t + \psi_t dB_t
\end{equation}
Then from (\ref{self}) it follows that
\begin{equation}\label{E.Z}
 dE_t = \phi_t dZ_t
\end{equation}
where
\begin{eqnarray}
 &&E_t = B_t^{-1} V_t\label{E}\\
 &&Z_t = B_t^{-1}S_t\label{Z}
\end{eqnarray}
So, (\ref{E.Z}) relates the discounted claim price $E_t$ to the discounted tradable price $Z_t$.

{}Let us now assume that we can construct a measure ${\bf Q}$ under which $Z_t$ is a martingale. Then we can construct a self-financing strategy which replicates the claim $X_T$ by setting
\begin{equation}\label{E.exp}
 E_t = \mathbb{E}\left(B_T^{-1} X_T\right)_{{\bf Q},{\cal F}_t}
\end{equation}
We then have $V_T = B_T E_T = X_T$. Since both $E_t$ and $Z_t$ are ${\bf Q}$-martingales, pursuant to the martingale representation theorem, $\phi_t$ is a previsible process. Furthermore, from (\ref{V}), (\ref{E}) and (\ref{Z}) we have
\begin{equation}\label{psi.bond}
 \psi_t = E_t - \phi_t Z_t
\end{equation}
so $\psi_t$ is also previsible. Continuity of $S_t$ and $B_t$ is key to previsibility of $\phi_t$ and $\psi_t$.

\subsubsection{A ``Caveat"}

{}However, if we think about relativistic Brownian motion in terms of a stochastic process, then it is discontinuous. By virtue of (\ref{P.corr}) it is a L{\'e}vy process, and the only continuous non-deterministic L{\'e}vy process is non-relativistic Brownian motion (with a drift). (See, e.g., \cite{Papa}.) I.e., the stochastic process corresponding to relativistic Brownian motion has jumps -- in fact, it is a pure jump process. Previsibility and replication are gone, so the above argument leading to the pricing formula (\ref{claim.prc}) no longer applies. Simply put, the market is incomplete and there is no perfect hedge. (See, e.g., \cite{Merton1}, \cite{Protter}.)

{}A simple way to see that there are jumps is to look at the transition density (\ref{P_K1}). As $t\rightarrow 0$, $t^{-1}{\widetilde P}(x,t)$ does not vanish for $x\neq 0$, so there are jumps\footnote{\, The second moment for both $t^{-1}{\widetilde P}(x,t)$ and ${\widehat P}(x) = t^{-1}{\widetilde P}(x,t)|_{t\rightarrow 0} = (c^2/\pi\sigma^2|x|) K_1(c|x|/\sigma^2)$ (which corresponds to the L\'evy measure (see, e.g., \cite{Papa}) for the jumps) is $\sigma^2$.} (cf. (\ref{gauss}), where we have no jumps). This is the case for any Hamiltonian ${\widehat H}$ other than the quadratic Hamiltonian for non-relativistic Brownian motion (and, more generally, any L{\'e}vy process other than non-relativistic Brownian motion with a drift). However, not all is lost. We can still require the self-financing property. The above argument still holds through (\ref{Z}). But we cannot apply the martingale representation theorem.

{}Instead, we can proceed as follows. First, the analog of It\^{o}'s lemma now reads:
\begin{equation}
 df(x,t) = {\widehat D} f(x,t)~dt + \sigma\partial_x f(x,t)~dx
\end{equation}
where the (generally, nonlocal) differential operator
\begin{equation}
 {\widehat D} = \partial_t - {\widetilde {\cal H}}(-\partial_x^2)
\end{equation}
Here, using our notations above, ${\widetilde {\cal H}}(u) = {\cal H}(u)/\hbar$. So, in the non-relativistic case, where we have ${\widetilde {\cal H}}(u) = mu/2\hbar = \sigma^2 u/2$, we recover the familiar form of It\^{o}'s lemma with\footnote{\, This is a local operator.} ${\widehat D} = \partial_t + {\sigma^2\over 2}\partial_x^2$. Next, as above, let us take the discounted stock process $Z_t$ to be a martingale: $Z_t = Z_0\exp(\zeta x + {\widetilde {\cal H}}(-\zeta^2) t)$. Then we have $dZ_t = \sigma\zeta Z_t dx$ and, by virtue of (\ref{E.Z}), we have the following differential equation for $E_t$:
\begin{equation}\label{D.E}
 {\widehat D} E_t = 0
\end{equation}
I.e., $E_t$ is a martingale (see below). The terminal condition is $E_T = B_T^{-1} X_T$. The solution to (\ref{D.E}) subject to this terminal condition is given by (\ref{E.exp}). More precisely, we can write (\ref{FP}) as ${\widehat D}_* {\widetilde P}(x,t) = 0$, where ${\widehat D}_* = \partial_t + {\widetilde {\cal H}}(-\partial_x^2)$. The general solution is given by
\begin{equation}
 {\widetilde P}(x,t) = \int {dk\over 2\pi} \exp(ikx)\exp\left(-t{\widetilde {\cal H}}(k^2)\right)\Omega(k)
\end{equation}
Above we required the tower law (semigroup property) to fix $\Omega(k)\equiv 1$. With this requirement, we have (\ref{E.exp}), i.e. (here $x^\prime$ corresponds to time $T$),
\begin{equation}
 E_t = \int_{-\infty}^\infty dx^\prime~{\widetilde P}\left(x^\prime - x, T - t\right) B_T^{-1} X_T
\end{equation}
Note that for $\Omega(k)\equiv 1$ we have $\lim_{t\uparrow T}{\widetilde P}(x^\prime - x, T - t) = \delta(x^\prime - x)$, so the terminal condition $E_T = B_T^{-1} X_T$ is satisfied. Without $\Omega(k)\equiv 1$ the terminal value pricing would not be possible. This is why the tower law is crucial. In fact, we can intuitively understand why as follows. The tower law simply implies that it makes no difference if we price a claim directly by starting at time $t=t_0$ and ending at time $t_f = T$, or if we price it in two steps by inserting an arbitrary intermediate time $t_0 < t_1 < t_f$, or multiple such times $t_0 < t_1 < \dots < t_{N-1} < t_N = t_f$, i.e., if we skeletonize our computation. The large $N$ limit of the skeletonized expectation value is simply Euclidean path integral. That is, the tower law is simply the semigroup property of path integral. From a physics viewpoint, we can think about the transition density ${\widetilde P}(x_f - x_0, t_f - t_0)$ as the propagator and we should be able to skeletonize it, or else we would have path dependence. There should be no path dependence in pricing the claim $X_T$, which is by assumption path-independent and only depends on $S_T$.

{}So, using the above approach, we can price claims, but this does not mean that we have a perfect hedge. If all paths were continuous, then the $(\phi_t,\psi_t)$ portfolio defined above would provide a perfect replicating hedge.\footnote{\, For the sake of simplicity, let us assume that the cash bond is deterministic. Then from (\ref{self}) it follows that $\phi_t = \partial V_t/\partial S_t$. Thus, for a constant risk-free interest rate $r$ we have $\phi_t = \exp\left(-r[T-t]\right)S_*\int_{x_*}^\infty dx^\prime~{\widetilde P}(x^\prime, T-t)~\exp(\zeta[x^\prime - x_*]) > 0$, which is finite (so $E_t$ is a martingale).} However, in the stochastic description we have jumps. So, how should one go about hedging? One can simply take a position that these jumps are an artifact of using a non-Gaussian distribution. If real-life prices have no jumps,\footnote{\, Real-life prices are discrete as there are minimum price increments, and real-life trading is not continuous. However, this discretization is not what we refer to above as in the continuum limit any residual hedging risk created therefrom goes to zero \cite{MertonSam}. What we have in mind are substantial jumps in prices unrelated to aforesaid discretization, to wit, due to unexpected information arriving in the market (earnings surprises, takeover bids, distress, etc.).} then we can hedge via the $(\phi_t,\psi_t)$ portfolio, and -- assuming the probability distribution we use is a good approximation of the real-life distribution -- this should do. However, if there are jumps in real-life prices, we cannot have a perfect hedge (assuming the market is incomplete) this way and we can resort to the well-studied risk diversification methodology of \cite{Merton1}.\footnote{\, Also see \cite{Cont} and references therein.}

\subsection{Quantum Field Theory?}

{}However, there is something unsettling about jumps in our context -- from a physics intuition viewpoint, that is. It makes little sense that we have jumps and here is why. We started from non-relativistic Brownian motion, which is simply Euclidean quantum mechanics, a well-defined theory. We extended it relativistically, which is certainly a well-defined thing to do. Yet we arrived at a pure jump model, with no continuous paths. Is a Euclidean quantum relativistic particle inconsistent or...?

{}The answer is perhaps unexpected from a finance viewpoint -- considering that jump processes have been utilized in the pricing context for decades now -- but has been well-known in physics for some decades longer... The description of a relativistic particle via quantum mechanics (which is a first-quantized theory) is simply inadequate. The correct description is in terms of a {\em quantum field theory} (i.e., a second-quantized theory, \cite{Dirac}), which has an infinite number of degrees of freedom. A simple way to understand this is to note that the relativistic Hamiltonian ${\widehat H}$ is nonlocal, i.e., it is an infinite series of spatial derivatives. What this means is that one should look for a local description via other degrees of freedom, and their number is infinite. The description for a relativistic particle is known. It is simply a (Euclidean) free massive scalar theory in $D$ dimensions (in our case $D=2$).\footnote{\, If we have $d$ Brownian motions, then $D=d+1$, $d$ spatial dimensions plus Euclidean time.} The scalar field is $\phi(x^\mu)$ (here, as before, $x^1 = x$ and $x^2 = ct$) and its action is given by\footnote{\, Note that this is not the same as quantum field theory of \cite{Baaquie1, Baaquie2}. Nor is our proposal related to the ``gauge theory of arbitrage" (GTA) of \cite{Ilinski1}, \cite{Ilinski2} -- for a popular discussion of GTA, see \cite{Dunbar}; for a critique of GTA, see \cite{Sornette}.}
\begin{equation}\label{scalar}
 S[\phi(x)] = \int d^2x~L(\phi,\partial_\mu\phi) = \int d^2x\left[\delta^{\mu\nu}\partial_\mu\phi\partial_\nu\phi + m^2c^2\phi^2/\hbar^2\right]
\end{equation}
Once this field theory is quantized, we have creation and annihilation operators, a Fock space of states which they create from the vacuum, etc. It is outside of the scope of this paper to make a precise connection between this field theory and finance. Here we simply note that the scalar field theory (\ref{scalar}) is perfectly local and description therethrough is continuous. The action (\ref{scalar}) can be augmented by nonlinear (local) interaction terms. Also, note that while here we focus on the relativistic Brownian motion case, one can consider elevating other nonlocal Hamiltonians to field theories as well.\footnote{\, For a proposal to extend {\em non-relativistic} Schr\"odinger action by adding nonlinear terms with a potential application to finance, see \cite{Nakayama}. Our proposal here is in the context of relativistic theories. However, the general idea is also applicable to non-relativistic cases as well.} Again, this is outside of the scope hereof and will be discussed elsewhere.

\section{Concluding Remarks}\label{sec.5}

{}The pitfall of the ``Gaussian approximation" arises in physics too. Thus, in usual (not Euclidean) time, consider the following Hamiltonian: $H = m^2c^4/\sqrt{m^2c^4 + p^2}$, where $p$ is the momentum, while $m$ and $c$ have the dimensions of mass and velocity, respectively, albeit $c$ is not necessarily  interpreted as the speed of light. If we expand this Hamiltonian, which is expressly positive-definite, in powers of $p^2$ and truncate it, we will get $H = mc^2 - p^2/2m$, which is not bounded from below and has a wrong-signed kinetic term -- such a state is known in physics as a ghost. However, this is merely an artifact of the ``Gaussian approximation" and the original (full, untruncated) theory a priori is well-defined. In fact, such a Hamiltonian arises \cite{dS,Massive} in massive gravity via gravitational Higgs mechanism \cite{Higgs}, \cite{thooft}. That is, often the ``Gaussian approximation" is not as innocuous as it might appear.\footnote{\, A flipside of this is the ``black swan" theme of \cite{Taleb}.}

{}In this paper we discuss an example of this in the financial setting. We consider a relativistic extension of Brownian motion. In fact, there is some literature with attempts to make Brownian motion relativistic, including in the financial context. However, curiously, most of these attempts appear to be focused on considering relativistic diffusion in Minkowski time. Yet, Minkowski time is utterly irrelevant in the financial context. Time in finance -- in physics language -- is Euclidean time, not Minkowski. There is no ``speed limit" analogous to the speed of light in the financial context. Our analog of the speed of light $c$ is not a ``speed limit". It is simply the characteristic speed of the diffusion at which relativistic effects become important. The spacetime signature (where the spatial direction $x$ corresponds to the variable describing the Brownian motion) is Euclidean, so in the zeroth approximation we have the Euclidean $SO(2)$ (not the Minkowski $SO(1,1)$) rotational invariance, where space and time are indistinguishable. However, as we discuss above, in the financial context such invariance ultimately must be broken and the relativistic effects manifest themselves via higher-order contributions to the Hamiltonian (and probability density), which result in a much softer asymptotic behavior at large $x$, i.e., fat tails. So, in this context we can think of fat tails as a relativistic effect, whereby at diffusion speeds around and above $c$, higher derivative (in fact, nonlocal) terms in the Hamiltonian simply cannot be ignored, i.e., the Gaussian approximation breaks down. One of the important points in our discussion is that, a priori one can construct an infinite family of probability densities, including the na\"{i}ve $SO(2)$ invariant density (which is the 2-dimensional Euclidean propagator one would use in physics), but it is utterly inapplicable in the financial context as it does not satisfy the tower law. The tower law and the requirement that the total probability be normalized to 1 then uniquely fix the probability density using an appropriately tweaked relativistic Hamiltonian. Martingales and option pricing then follow, albeit there is a subtlety with martingales, to wit, there is an upper bound on log-volatility due to the fact that the probability density decays only exponentially at large values of $|x|$ (up to a prefactor proportional to a power of $|x|$), not as a Gaussian density.

{}In terms of practical applications, the relativistic setting we discuss here provides a 1-parameter extension of the BSM model, whereby the probability density is automatically fat-tailed, hence volatility smiles. There exist other extensions (e.g., stochastic volatility and local volatility models). In contrast, in ours all parameters are constant and the relativistic interpretation is appealing. However, we emphasize that in principle we can have other (myriad) choices of the Hamiltonian in the Fokker-Planck-like equation (\ref{FP}) and other interesting models can be constructed. For a recent example, see, e.g., \cite{ACKK}, and also references therein.

{}Let us note that in \cite{Romero} a second-order (in time derivatives) equation for a probability distribution is attempted to be constructed by starting from the massive Klein-Gordon equation, Wick-rotating it and ad hoc inserting a multiplicative factor that appears to be related to a martingale for the Gaussian probability distribution. There are two issues with this approach. First, as we discuss above, the differential equation for the probability distribution must be first-order (in time derivatives) or else it is unclear how to construct martingales and price claims. Second, basing the aforesaid ad hoc multiplicative factor on a Gaussian (non-relativistic) martingale is inconsistent with the relativistic probability distribution. Put differently, it is unclear how to price claims in the setting of \cite{Romero}. Not surprisingly, their probability distribution differs from ours, and as we argue above, ours is uniquely fixed by requiring proper normalization and the tower law\footnote{\, Which conditions the distribution of \cite{Romero} does not satisfy.} required for self-financing strategies.

{}Finally, perhaps the most intriguing conclusion hereof is that, as we argue above, a nonlocal stochastic description of relativistic Brownian motion (and other such L{\'e}vy processes) is inadequate -- a local quantum field theory description is dictated by physics considerations. The infinite number of degrees of freedom associated with a quantum field theory description would bode well with market incompleteness and hedging techniques such as in \cite{Merton1}, \cite{Cont}. Time will tell...

\subsection*{Acknowledgments}
{}I would like to thank Peter Carr for enlightening discussions. I would also like to thank anonymous referees for comments and suggesting additional references.

\end{document}